\documentstyle [12pt] {article}
\begin{document}
\setcounter{page}{1}
\def\theequation{\arabic{section}.\arabic{equation}}
\def\theequation{\thesection.\arabic{equation}}
\setcounter{section}{0}

\title{Effective quark model with chiral $U(3)\times U(3)$ symmetry for
baryon octet and decuplet}

\author{A. N. Ivanov\thanks{E--mail: ivanov@kph.tuwien.ac.at, Tel.:
+43--1--58801--5598, Fax: +43--1--5864203}~${\textstyle ^\ddagger}$ , M.
Nagy\thanks{E--mail: fyzinami@savba.sk, Institute of Physics, Slovak
Academy of Sciences, D$\acute{\rm u}$bravsk$\acute{\rm a}$ cesta 9, SK--842
28 Bratislava, Slovakia}  ~and~  N. I. Troitskaya\thanks{Permanent Address:
State Technical University, Department of Nuclear
Physics, 195251 St. Petersburg, Russian Federation}}

\date{}

\maketitle

\begin{center}
{\it Institut f\"ur Kernphysik, Technische Universit\"at Wien, \\
Wiedner Hauptstr. 8-10, A-1040 Vienna, Austria}
\end{center}

\vskip1.0truecm
\begin{center}
\begin{abstract}
We suggest an effective quark model for low--lying baryon octet and
decuplet motivated by QCD with a linearly rising confinement potential
incorporating the extended Nambu--Jona--Lasinio (ENJL) model with linear
realization of chiral $U(3)\times U(3)$ symmetry. Baryons are considered as
external heavy states coupled to local three--quark currents with fixed
spinorial structure and to low--lying mesons through quark--meson
interactions defined in the ENJL--model. In the constituent quark loop
representation we have calculated the coupling constants of the ${\rm \pi N
N}$, ${\rm \pi N \Delta}$ and ${\rm \gamma N \Delta}$ interactions and the
$\sigma_{\rm \pi N}$--term. The obtained results are in reasonable
agreement with experimental data and other effective field theory
approaches.
\end{abstract}
\end{center}
\vspace{0.2in}

\begin{center}
PACS number(s): 11.30.Hv, 11.30.Rd, 12.38.Aw, 12.39.Ki, 12.39.Fe,\\
13.40.Hq, 13.75.Gx, 14.20.Dh, 14.20.Jn
\end{center}

\newpage

\section{Introduction}
\setcounter{equation}{0}

It is well--known that low--energy properties of low--lying mesons and
baryons can be suitably described by phenomenological chiral Lagrangians in
the tree approximation [1--4]. The step beyond the tree approximation has
been done by Weinberg [5] and then systematized by Gasser and Leutwyler
within Chiral perturbation theory (CHPT) [6]. The application of CHPT to
the description of low--lying baryons coupled to low--lying mesons showed
[7,8] the necessity of the consideration of the baryons as very heavy
states  and the development of  large $M_B$, the baryon mass, expansion.
The numerous applications of this approach have been done by Bernard et al.
and collected in the review [9].

The success of CHPT had inspired the development of different approaches to
the description of low--energy interactions of low--lying hadrons based on
the effective chiral Lagrangians derived within the extended
Nambu--Jona--Lasinio (ENJL) model [10] motivated by QCD with a non--linear
[11] and linear [12] realization of chiral $U(3)\times U(3)$ symmetry. The
relation of the ENJL model with linear realization of chiral $U(3)\times
U(3)$ symmetry to QCD with linearly rising confinement potential has been
shown in [13]. The properties of octet and decuplet of low--lying baryons
within QCD with linearly rising confinement potential have been
investigated in [14]. There has been shown that these baryons are only
three--quark states [15], this means that they do not contain the
contributions of bound diquark states and so on, the spinorial structure of
quark currents of which is defined as products of axial--vector diquark
densities and a quark field transforming under $SU(3)_f\times SU(3)_c$
group like $(\underline{6}_f,\tilde{\underline{3}}_c)$ and
$(\underline{3}_f,\underline{3}_c)$ multiplets, respectively. The former
agrees with the structure of baryon quark currents suggested by Ioffe [16],
Pascual and Tarrach [17], and Reinders, Rubinstein and Yazaki [18].

This paper is to apply the dynamical constraints on the structure of
baryons obtained in Ref.[14] to the construction of an effective quark
model for the description of low--energy interactions of the baryon octet
and decuplet coupled to low--lying mesons. According to these dynamical
constraints imposed by a linearly rising interquark confinement potential
[14] the spinorial structure of the three--quark currents should be defined
by the products of axial--vector diquark densities and a quark field
transforming under $SU(3)_f\times SU(3)_c$ group like
$(\underline{6}_f,\tilde{\underline{3}}_c)$ and
$(\underline{3}_f,\underline{3}_c)$ multiplets, respectively. This allows
to construct the three--quark currents with quantum numbers of the baryon
octet and decuplet in the following form [14--18]:

\begin{eqnarray}\label{label1.1}
\hspace{-0.3in}\eta_{\rm N}(x) &=& - \,\varepsilon^{ijk}\,[\bar{u^c}_i(x)
\gamma^{\mu} u_j(x)]\gamma_{\mu}\gamma^5 d_k(x)\,,\nonumber\\
\hspace{-0.3in}\eta_{\rm \Sigma}(x) &=& \varepsilon^{ijk}\,[\bar{u^c}_i(x)
\gamma^{\mu} u_j(x)]\gamma_{\mu}\gamma^5 s_k(x)\,,\nonumber\\
\hspace{-0.3in}\eta_{\rm \Lambda}(x) &=& - \sqrt{\frac{2}{3}}
\varepsilon^{ijk} \{[\bar{u^c}_i(x) \gamma^{\mu}
s_j(x)]\gamma_{\mu}\gamma^5 d_k(x) - [\bar{d^c}_i(x) \gamma^{\mu}
s_j(x)]\gamma_{\mu}\gamma^5 u_k(x)\},\nonumber\\
\hspace{-0.3in}\eta_{\rm \Xi}(x) &=&  \varepsilon^{ijk}\,[\bar{s^c}_i(x)
\gamma^{\mu} s_j(x)]\gamma_{\mu}\gamma^5 u_k(x),
\end{eqnarray}
\begin{eqnarray}\label{label1.2}
\hspace{-0.3in}\eta^{\mu}_{\rm \Delta}(x) &=&
\varepsilon^{ijk}\,[\bar{u^c}_i(x) \gamma^{\mu} u_j(x)] u_k(x)\,,\nonumber\\
\hspace{-0.3in}\eta^{\mu}_{\rm \Sigma^*}(x) &=&
\sqrt{\frac{1}{3}}\,\varepsilon^{ijk}\,\{2\,[\bar{u^c}_i(x) \gamma^{\mu}
s_j(x)] u_k(x) + [\bar{u^c}_i(x) \gamma^{\mu} u_j(x)] s_k(x)\}\,,\nonumber\\
\hspace{-0.3in}\eta^{\mu}_{\rm \Xi^*}(x) &=&
\sqrt{\frac{1}{3}}\,\varepsilon^{ijk}\,\{2\,[\bar{s^c}_i(x) \gamma^{\mu}
u_j(x)] s_k(x) + [\bar{s^c}_i(x) \gamma^{\mu} s_j(x)] u_k(x)\}\,,\nonumber\\
\hspace{-0.3in}\eta^{\mu}_{\rm \Omega}(x) &=&
\varepsilon^{ijk}\,[\bar{s^c}_i(x) \gamma^{\mu} s_j(x)] s_k(x),
\end{eqnarray}
where $i,j$ and $k$ are colour indices, then $\bar{q^c}(x) = q(x)^T C$ and
$C = - C^T = - C^{\dagger}$ is a matrix of a charge conjugation, $T$ is a
transposition.

In our approach baryons are external heavy states [8,9] coupled to
three--quark currents defined by Eq.(\ref{label1.1}) and
Eq.(\ref{label1.2}) as
\begin{eqnarray}\label{label1.3}
{\cal L}_{\rm int}(x) = \frac{1}{\sqrt{2}}\,g_{\rm
B}\,\bar{p}(x)\,\eta_{\rm N}(x) + g_{\rm B}\,\bar{\Delta}^{++}_{\mu}(x)
\eta^{\mu}_{\Delta}(x) + \ldots + {\rm h.c.},
\end{eqnarray}
where $p(x)$ and $\Delta^{++}_{\mu}(x)$ are the fields of the proton and
the $\Delta$--resonance, respectively, the ellipses denote the contribution
of interactions of other components of octet and decuplet, $g_{\rm B}$ is a
phenomenological coupling constant, and a factor $1/\sqrt{2}$ is introduced
due to a non--relativistic quark model with $SU(6)$ symmetry [19]. We
should emphasize that in our consideration the $\Delta$--resonance and
other components of decuplet are the Rarita--Schwinger fields obeying the
subsidiary conditions
\begin{eqnarray}\label{label1.4}
\partial^{\mu}\Delta_{\mu}(x) = \gamma^{\mu}\Delta_{\mu}(x) = 0\,,\quad
{\rm etc}.
\end{eqnarray}
To low--lying mesons baryons couple through quark--meson interactions
induced within the ENJL--model with linear realization of chiral
$U(3)\times U(3)$ symmetry [12,13].

The paper is organized as follows. In Section 2 we calculate the $g_{\rm
\pi NN}$ and $g_{\rm \pi N\Delta}$ coupling constants. We find the ratio
$g_{\rm \pi N\Delta}/g_{\rm \pi NN}=2$ in agreement with both experimental
data and other approaches [9]. In Section 3 we calculate the coupling
constant $g_{\rm \gamma N \Delta}$ of the $\Delta \to N + \gamma$ decays.
In Section 4 we calculate chiral corrections to the mass of the proton
related to the $\sigma_{\rm \pi N}$--term of the ${\rm \pi N}$--scattering
and estimate the value of the $\sigma_{\rm \pi N}$--term. In Conclusion we
discuss the obtained results.

\section{$g_{\rm \pi NN}$ and $g_{\rm \pi N\Delta}$ coupling constants}
\setcounter{equation}{0}

The coupling constant $g_{\rm \pi NN}$ of the ${\rm \pi NN}$--interaction
we define by a phenomenological ${\rm \pi^0 pp}$--interaction [20]:
\begin{eqnarray}\label{label2.1}
{\cal L}^{\rm \pi^0 pp}_{\rm eff}(x) = g_{\rm \pi NN}\,[\bar{p}(x)
i\gamma^5 p(x)]\,\pi^0(x),
\end{eqnarray}
where $\pi^0(x)$ is the $\pi^0$--meson field. The pion fields  couple to
current quark fields through the interactions [12]:
\begin{eqnarray}\label{label2.2}
\hspace{-0.2in}{\cal L}^{\rm \pi^0 qq}(x) &=& \frac{g_{\rm \pi
qq}}{\sqrt{2}}\,[\bar{u}(x) i\gamma^5 u(x) - \bar{d}(x) i\gamma^5
d(x)]\,\pi^0(x) \nonumber\\
&&+ g_{\rm \pi qq}\,[\bar{u}(x) i\gamma^5 d(x)]\,\pi^+(x) + {\rm h.c.}
\end{eqnarray}
The quark--meson coupling constant $g_{\rm \pi qq} = \sqrt{2}m/F_0$ [12] is
given in terms of the constituent quark mass $m = 0.33\,{\rm GeV}$ and the
PCAC constant $F_0 = 0.092\,{\rm GeV}$ calculated in the chiral limit [12].
The effective Lagrangian ${\cal L}^{\rm \pi^0 pp}_{\rm eff}(x)$ defined by
the interactions Eq.(\ref{label1.3}) and Eq.(\ref{label2.2}) reads
\begin{eqnarray}\label{label2.3}
\hspace{-0.4in}&&\int d^4x\,{\cal L}^{\rm \pi^0 pp}_{\rm eff}(x) = -
\Bigg(\frac{g_{\rm \pi qq}}{\sqrt{2}}\Bigg)\,\Bigg(\frac{g^2_{\rm
B}}{2}\Bigg)\int d^4x d^4x_1 d^4x_2 \pi^0(x_2) \nonumber\\
\hspace{-0.4in}&&\bar{p}(x)<0|{\rm T}(\eta_{\rm N}(x)[\bar{u}(x_2)
i\gamma^5 u(x_2) - \bar{d}(x_2) i\gamma^5 d(x_2)]\bar{\eta}_{\rm
N}(x_1))|0> p(x_1) ,
\end{eqnarray}
where T is a time--ordering operator, $\bar{\eta}_{\rm N}(x) =
\varepsilon^{ijk}\,\bar{d}_i(x)\gamma_{\nu}\gamma^5 [\bar{u}_j(x)
\gamma^{\nu} u^c_k(x)]$ and $u^c(x) = C\bar{u}(x)^T$.

By applying formulae of quark conversion [12] (Ivanov) we can represent the
r.h.s. of Eq.(\ref{label2.3}) in terms of constituent quark diagrams. In
the momentum representation they read
\begin{eqnarray}\label{label2.4}
\hspace{-0.4in}&&\int d^4x\,{\cal L}^{\rm \pi^0 pp}_{\rm eff}(x) =\nonumber\\
\hspace{-0.4in}&&=- \frac{3}{2} \Bigg(\frac{g_{\rm \pi qq}}{\sqrt{2}}\Bigg)
\Bigg(\frac{g_{\rm B}}{8\pi^2}\Bigg)^2\int d^4x \int \frac{d^4x_1
d^4p_1}{(2\pi)^4}\frac{d^4x_2 d^4p_2}{(2\pi)^4} e^{-i(x - x_1)\cdot p_1}
e^{-i(x - x_2)\cdot p_2} \pi^0(x_2)\nonumber\\
\hspace{-0.4in}&&\int \frac{d^4k_1 }{\pi^2i}\frac{d^4k_2}{\pi^2i}
\bar{p}(x) \Bigg[\gamma^{\mu}\gamma^5 \frac{1}{m - \hat{k}_1 - \hat{k}_2 -
\hat{p}_1 - \hat{p}_2}i\gamma^5 \frac{1}{m - \hat{k}_1 - \hat{k}_2 -
\hat{p}_1}\gamma^{\nu}\gamma^5\Bigg]p(x_1)\nonumber\\
\hspace{-0.5in}&&{\rm tr}\Bigg\{\frac{1}{m +
\hat{k}_1}\gamma_{\nu}\frac{1}{m - \hat{k}_2}\gamma_{\mu}\Bigg\}.
\end{eqnarray}
The calculation of the momentum integrals representing constituent quark
diagrams should be carried out keeping only the divergent parts and
dropping the contributions of the parts finite in the infinite limit of the
cut--off. Such a prescription realizes a naive description of the quark
confinement. Indeed, dropping the finite parts of quark diagrams we remove
the imaginary parts of them and suppress by this the appearance of quarks
in the intermediate states of low--energy hadron interactions. This naive
description of confinement has turned out to be rather useful for the
derivation of Effective Chiral Lagrangians [10--12]. As has been shown in
Ref. [13] this prescription can be justified in QCD with a linearly rising
interquark confinement potential. Thereby, within such a naive approach to
the quark  confinement mechanism one can bridge  quantitatively the quark
and the hadron level of the description of strong low--energy interactions
of hadrons. The cut--off $\Lambda_{\chi}=0.94\,{\rm GeV}$ and the
constituent quark mass $m=0.33\,{\rm GeV}$ [12] can be considered in such
an approach as input parameters and fixed at one--loop approximation via
the experimental values of the $\rho\pi\pi$ coupling constant $g_{\rho}$
and the leptonic pion constant $F_{\pi}$ [12]. Thus, we should accentuate
that in the effective quark models based on the NJL approach, or
equivalently QCD with a linearly rising interquark confinement potential
[13], quark diagrams lose the meaning of quantum field theory objects and
only display how quark flavours can be transferred form an initial hadron
state to a final hadron state in hadron--hadron low--energy transitions.
The coupling constants of such transitions, described in terms of divergent
parts of quark diagrams and depending of the cut--off
$\Lambda_{\chi}=0.94\,{\rm GeV}$ and the constituent quark mass
$m=0.33\,{\rm GeV}$, can be expressed in terms of effective
phenomenological coupling constants of low--energy hadron interactions
given by Effective Chiral Lagrangians [4].

For the computation of the momentum integrals we assume following Jenkins
and Manohar [8] that the proton is a very heavy state and its mass is much
larger than other momenta in the integrand such as $p^2_1 = M^2_{\rm N}$,
where $M_{\rm N}$ is an averaged  mass of the baryon octet. For numerical
estimates we set below $M_{\rm N}=940\,{\rm MeV}$. In this picture a very
heavy source (the proton) is surrounded by a cloud of light (almost
massless) particles [9]. This resembles completely the methods used in the
Heavy Quark Effective theory (HQET) [21].

Thus, keeping the leading order in large $M_{\rm N}$ expansion we reduce
the r.h.s. of Eq.(\ref{label2.4}) to the form
\begin{eqnarray}\label{label2.5}
\hspace{-0.4in}&&\int d^4x\,{\cal L}^{\rm \pi^0 pp}_{\rm eff}(x) = \nonumber\\
\hspace{-0.4in}&&=- \frac{3}{2} \Bigg(\frac{g_{\rm \pi qq}}{\sqrt{2}}\Bigg)
\Bigg(\frac{g_{\rm B}}{8\pi^2}\Bigg)^2\int \! d^4x\! \int \frac{d^4x_1
d^4p_1}{(2\pi)^4}\frac{d^4x_2 d^4p_2}{(2\pi)^4} e^{-i(x - x_1)\cdot p_1}
e^{-i(x - x_2)\cdot p_2}\pi^0(x_2) \nonumber\\
\hspace{-0.4in}&&\int \frac{d^4k_1 }{\pi^2i}\frac{d^4k_2}{\pi^2i}
\,\bar{p}(x) \Bigg[\gamma^{\mu}\gamma^5 \frac{\hat{p}_1}{M^2_{\rm N}}i
\gamma^5 \frac{\hat{p}_1}{M^2_{\rm N}}\gamma^{\nu}\gamma^5\Bigg] p(x_1)
{\rm tr}\Bigg\{\frac{1}{m + \hat{k}_1}\gamma_{\nu}\frac{1}{m -
\hat{k}_2}\gamma_{\mu}\Bigg\}.
\end{eqnarray}
The replacement of the constituent quark Green function
\begin{eqnarray}\label{label2.6}
\frac{1}{m - \hat{k}_1 - \hat{k}_2 - \hat{p}_1} \to -
\frac{\hat{p}_1}{M^2_{\rm N}}
\end{eqnarray}
agrees with the heavy baryon [8,9] and HQET [21] approaches. Indeed, in
accordance with [8,9] and HQET [21] we obtain
\begin{eqnarray}\label{label2.7}
&&\frac{1}{m - \hat{k}_1 - \hat{k}_2 - \hat{p}_1} = \frac{m + \hat{k}_1 +
\hat{k}_2 + \hat{p}_1}{m^2 - (k_1 + k_2 + p_1)^2 - i0} = \nonumber\\
&&= - \frac{m + \hat{k}_1 + \hat{k}_2 + \hat{p}_1}{M^2_{\rm N} + 2(k_1 +
k_2)\cdot p_1 - m^2 + (k_1 + k_2)^2 + i0}=\nonumber\\
&&=- \frac{1}{M_{\rm N}}\,\frac{\displaystyle \frac{m + \hat{k}_1 +
\hat{k}_2}{M_{\rm N}} + \hat{v}_1}{\displaystyle 1 + \frac{2(k_1 +
k_2)\cdot v_1}{M_{\rm N}} - \frac{m^2}{M^2_{\rm N}} + \frac{(k_1 +
k_2)^2}{M^2_{\rm N}} + i0},
\end{eqnarray}
where we have set $p^{\mu}_1 = M_{\rm N}\,v^{\mu}_1$ [8,9,21]. In the case
of $m = M_{\rm N}$ and in the limit  $M_{\rm N} \to \infty$ [8,9,21] we
arrive at the well--known expression for the Green function of a heavy
baryon (or a heavy quark in HQET [21]) [8,9]:
\begin{eqnarray}\label{label2.8}
\frac{1}{m - \hat{k}_1 - \hat{k}_2 - \hat{p}_1} &=&\frac{1}{M_{\rm N} -
\hat{k}_1 - \hat{k}_2 - M_{\rm N}\hat{v}_1} \to \nonumber\\
&\to& - \Bigg(\frac{1 + \hat{v}_1}{2}\Bigg)\,\frac{1}{(k_1 + k_2)\cdot v_1
+ i0}.
\end{eqnarray}
In our case $m\ll M_{\rm N}$, therefore, in the limit $M_{\rm N} \to
\infty$ [8,9] we arrive at the expression Eq.(\ref{label2.6}).

Let us discuss the choice of variables in the momentum integrals of
Eq.(\ref{label2.4}). These integrals represent the constituent quark
diagrams with three virtual quark lines two of which are confined by the
trace and the third is joined to the initial and the final proton fields.
For convenience we would call this third quark a spectator. The main
problem of the computation of the momentum integrals like those of
Eq.(\ref{label2.4}) is in the overlap of the virtual momenta $k_1$ and
$k_2$. The former makes rather complicated the computation of the momentum
integrals. Therefore, the choice of variables should help to disconnect the
momenta $k_1$ and $k_2$ in the limit $M_{\rm N} \to \infty$. This can be
reached assuming that the proton momentum transfers itself from the initial
state to the final one by the spectator.  This factorizes the momentum
integrals over $k_1$ and $k_2$. For the sake of self--consistency of the
approach we should use this choice of the virtual momenta, when the
momentum of the initial baryon transfers itself to the final one by the
spectator, for any application of the model to the description of
low--energy interactions of the baryon octet and decuplet.

The result of the integration over momenta $k_1$ and $k_2$ can be
represented in terms of the quark condensate [12]
\begin{eqnarray}\label{label2.9}
\int \frac{d^4k_1 }{\pi^2i}\frac{d^4k_2}{\pi^2i} {\rm tr}\Bigg\{\frac{1}{m
+ \hat{k}_1}\gamma_{\nu}\frac{1}{m - \hat{k}_2}\gamma_{\mu}\Bigg\} =
\Bigg(\frac{8\pi^2}{3}\Bigg)^2\,<\bar{q}q>^2\,g_{\mu\nu}.
\end{eqnarray}
After some algebra of Dirac matrices we obtain the effective Lagrangian $ $
in the form
\begin{eqnarray}\label{label2.10}
{\cal L}^{\rm \pi^0 p p}_{\rm eff}(x) &=&6\,\Bigg(\frac{g_{\rm \pi
qq}}{\sqrt{2}}\Bigg)\,\Bigg(\frac{g_{\rm
B}}{8\pi^2}\Bigg)^2\Bigg(\frac{8\pi^2}{3}\Bigg)^2\frac{<\bar{q}q>^2}{M^2_{\rm N}
}\,\,\bar{p}(x)\,i\gamma^5\,p(x)\,\pi^0(x)  = \nonumber\\
&=&\Bigg[g^2_{\rm
B}\,\frac{2}{3}\,\frac{m}{F_0}\,\frac{<\bar{q}q>^2}{M^2_{\rm
N}}\Bigg]\,\bar{p}(x)\,i\gamma^5\,p(x)\,\pi^0(x).
\end{eqnarray}
This yields the coupling constant $g_{\rm \pi NN}$:
\begin{eqnarray}\label{label2.11}
g_{\rm \pi NN} = g^2_{\rm
B}\,\frac{2}{3}\,\frac{m}{F_0}\,\frac{<\bar{q}q>^2}{M^2_{\rm N}}.
\end{eqnarray}
Now let us proceed to the computation of the coupling constant $g_{\rm \pi
N \Delta}$ of the $\pi {\rm N} \Delta$--interaction.

The most general form of the $\pi {\rm N} \Delta$--interaction compatible
with requirements of chiral symmetry reads [22]
\begin{eqnarray}\label{label2.12}
{\cal L}^{\rm \pi N \Delta}_{\rm eff}(x) = \frac{g_{\rm \pi N \Delta}}{2
M_N}\,\bar{\Delta}^a_{\mu}(x)\,\Theta^{\mu\nu}\,N(x)\,\partial_{\mu}\pi^a(x) + {
\rm h.c.} ,
\end{eqnarray}
where $\Delta^a_{\mu}(x)$ is the $\Delta$--resonance field, the isotopical
index $a$ runs over $a=1,2,3$,
\begin{eqnarray}\label{label2.13}
\begin{array}{llcl}
&&\Delta^1=\,{\frac{1}{\sqrt{2}}}\Biggr(\begin{array}{c}
\Delta^{++}-\Delta^0/\sqrt{3} \\ \Delta^+/\sqrt{3} - \Delta^-
\end{array}\Biggl)\,,\,
\Delta^2=\,\frac{i}{\sqrt{2}} \Biggr(\begin{array}{c}
\Delta^{++}+\Delta^0/\sqrt{3} \\ \Delta^+/\sqrt{3} + \Delta^-
\end{array}\Biggl),\\
&&\Delta^3=\,-\sqrt{\frac{2}{3}}\Biggr(\begin{array}{c} \Delta^+ \\
\Delta^0 \end{array}\Biggl)\,.
\end{array}
\end{eqnarray}
The nucleon field $N(x)$ is the isotopical doublet with components $N(x) =
(p(x), n(x))$. The tensor $\Theta^{\mu\nu}$ is given by [22]:
$\Theta^{\mu\nu} = g^{\mu\nu} -(Z + 1/2)\,\gamma^{\mu}\gamma^{\nu}$,
where the parameter $Z$ is fully arbitrary. It describes the ${\pi N
\Delta}$--interaction off--mass shell of the $\Delta$--resonance.  The
propagator of the $\Delta$--field is defined [23]:
\begin{eqnarray}\label{label2.14}
<0|{\rm
T}\Big(\Delta_{\mu}(x_1)\bar{\Delta}_{\nu}(x_2)\Big)|0>=-i\,S_{\mu\nu}(x_1-x_2).
\end{eqnarray}
In the momentum representation $S_{\mu\nu}(x)$ reads [23,24]:
\begin{eqnarray}\label{label2.15}
S_{\mu\nu}(p) = \frac{1}{M_{\Delta} - \hat{p}}\Bigg(- g_{\mu\nu} +
\frac{1}{3} \gamma_{\mu}\gamma_{\nu} +
\frac{1}{3}\frac{\gamma_{\mu}p_{\nu}-\gamma_{\nu}p_{\mu}}{M_{\Delta}} +
\frac{2}{3}\frac{p_{\mu}p_{\nu}}{M^2_{\Delta}}\Bigg),
\end{eqnarray}
where $M_{\Delta}$ is an averaged mass of the baryon decuplet. It can be
kept of order $M_{\rm N}$. The Green function of the free $\Delta$--field
Eq.(\ref{label2.14}) is related to the free Lagrangian given by [25]
\begin{eqnarray}\label{label2.16}
\hspace{-0.3in}{\cal L}^{\Delta}_{\rm kin}(x) =\bar{\Delta}_{\mu}(x)[
-(i\gamma^{\alpha}\partial_{\alpha} - M_{\Delta}) g^{\mu\nu} + \frac{1}{4}
\gamma^{\mu}\gamma^{\beta}(i\gamma^{\alpha}\partial_{\alpha} - M_{\Delta})
\gamma_{\beta}\gamma^{\nu}] \Delta_{\nu}(x).
\end{eqnarray}
In the component form the $\pi {\rm N}\Delta$--interaction
Eq.(\ref{label2.12}) reads
\begin{eqnarray}\label{label2.17}
{\cal L}^{\rm \pi N \Delta}_{\rm eff}(x) =\frac{g_{\rm \pi N \Delta}}{2
M_N}\,\bar{\Delta}^{++}_{\mu}(x) \Theta^{\mu\nu} p(x)
\partial_{\nu}\pi^+(x) + \ldots.
\end{eqnarray}
The effective Lagrangian ${\cal L}^{\rm \pi^+ p \Delta^{++}}_{\rm eff}(x)$
defined by Eq.(\ref{label1.3}) and Eq.(\ref{label2.2}) reads
\begin{eqnarray}\label{label2.18}
\hspace{-0.2in}&&\int d^4x\,{\cal L}^{\rm \pi^+ p \Delta^{++}}_{\rm eff}(x)
= - \Bigg(\frac{g_{\rm \pi qq}}{\sqrt{2}}\Bigg)\,g^2_{\rm B}\int d^4x
d^4x_1 d^4x_2 \,\nonumber\\
\hspace{-0.2in}&&\bar{\Delta}^{++}_{\mu}(x)\,<0|{\rm T}(\eta^{\mu}_{\rm
\Delta}(x)[\bar{u}(x_2) i\gamma^5 d(x_2)]\bar{\eta}_{\rm N}(x_1))|0>
p(x_1)\,\pi^+(x_2).
\end{eqnarray}
The computation of the $g_{\rm \pi N \Delta}$ coupling constant we perform
on--mass shell of the $\Delta$--resonance. The application of formulae of
quark conversion represents the r.h.s. of Eq.(\ref{label2.18}) in the form
of constituent quark diagrams determined by the momentum integrals as
follows:
\begin{eqnarray}\label{label2.19}
\hspace{-0.3in}&&\int d^4x\,{\cal L}^{\rm \pi^+ p \Delta^{++}}_{\rm eff}(x)
= \nonumber\\
\hspace{-0.3in}&&= 3 \Bigg(\frac{g_{\rm \pi qq}}{\sqrt{2}}\Bigg)
\Bigg(\frac{g_{\rm B}}{8\pi^2}\Bigg)^2\int d^4x \int \frac{d^4x_1
d^4p_1}{(2\pi)^4}\frac{d^4x_2 d^4p_2}{(2\pi)^4}\,e^{ -i(x - x_1)\cdot p_1}
e^{ -i(x - x_2)\cdot p_2}\nonumber\\
\hspace{-0.3in}&&\int \frac{d^4k_1 }{\pi^2i}\frac{d^4k_2}{\pi^2i}
\bar{\Delta}^{++}_{\mu}(x) \Bigg[\frac{1}{m - \hat{k}_1 - \hat{k}_2 -
\hat{p}_1 - \hat{p}_2}i\gamma^5 \frac{1}{m - \hat{k}_1 - \hat{k}_2 -
\hat{p}_1}\gamma_{\nu}\gamma^5\Bigg] p(x_1)\nonumber\\
\hspace{-0.3in}&&{\rm tr}\Bigg\{\frac{1}{m +
\hat{k}_1}\gamma^{\nu}\frac{1}{m -
\hat{k}_2}\gamma^{\mu}\Bigg\}\,\pi^+(x_2).
\end{eqnarray}
In the large $M_{\rm N}$ expansion due to the constraints
Eq.(\ref{label1.4}) the nontrivial contribution to the effective ${\rm \pi
N \Delta}$---interaction appears at first order in the pion momentum
expansion. This contribution reads
\begin{eqnarray}\label{label2.20}
\hspace{-0.2in}{\cal L}^{\rm \pi^+ p \Delta^{++}}_{\rm eff}(x) &=&6
\Bigg(\frac{g_{\rm \pi qq}}{\sqrt{2}}\Bigg) \Bigg(\frac{g_{\rm
B}}{8\pi^2}\Bigg)^2\Bigg(\frac{8\pi^2}{3}\Bigg)^2\frac{<\bar{q}q>^2}{M^3_{\rm N}
}\,\bar{\Delta}^{++}_{\mu}(x)\,p(x)\,\partial^{\mu}\pi^+(x) = \nonumber\\
&=&\frac{1}{2 M_{\rm N}}\,\Bigg[g^2_{\rm
B}\,\frac{4}{3}\,\frac{m}{F_0}\,\frac{<\bar{q}q>^2}{M^2_{\rm
N}}\Bigg]\,\bar{\Delta}^{++}_{\mu}(x)\,p(x)\,\partial^{\mu}\pi^+(x)
\end{eqnarray}
and gives the $g_{\rm \pi N\Delta}$ coupling constant:
\begin{eqnarray}\label{label2.21}
g_{\rm \pi N\Delta} = g^2_{\rm
B}\,\frac{4}{3}\,\frac{m}{F_0}\,\frac{<\bar{q}q>^2}{M^2_{\rm N}}.
\end{eqnarray}
Thus, in our approach the ratio of the coupling constants $g_{\rm \pi
N\Delta}$ and $g_{\rm \pi NN}$ is defined $g_{\rm \pi N\Delta}/g_{\rm \pi
NN}= 2$. This agrees with the experimental data and other effective field
theory approaches [9]. The parameter $Z$ is left undefined, since we have
computed $g_{\rm \pi N\Delta}$ on--mass shell of the $\Delta$--resonance,
while $Z$ describes the ${\rm \pi N\Delta}$--interaction off--mass shell.
\section{$g_{\rm \gamma N \Delta}$ coupling constant}
\setcounter{equation}{0}

Assuming that the transition $\Delta \to N + \gamma$ is primarily a
magnetic one [26,27] the effective Lagrangian describing the $\Delta \to N
+ \gamma$ decays can be defined on--mass shell of the $\Delta$--resonance
as [26,27]:
\begin{eqnarray}\label{label3.1}
{\cal L}^{\rm \gamma N \Delta}_{\rm eff}(x) &=& -\,i\, e\,\frac{g_{\rm
\gamma N \Delta}}{2M_{\rm
N}}\,\bar{\Delta}^3_{\mu}(x)\,\gamma_{\nu}\gamma^5 N(x)\,F^{\mu\nu}(x) +
{\rm h.c.}= \nonumber\\
&=& i\,e\,\sqrt{\frac{2}{3}}\,\frac{g_{\rm \gamma N \Delta}}{2M_{\rm
N}}\,\bar{\Delta}^+_{\mu}(x)\,\gamma_{\nu}\gamma^5 p(x)\,F^{\mu\nu}(x) +
\ldots + {\rm h.c.},
\end{eqnarray}
where $F^{\mu\nu}(x) = \partial^{\mu} A^{\nu}(x) - \partial^{\nu}
A^{\mu}(x)$ and $A^{\mu}(x)$ is the photon field, $e$ is the electric
charge of the proton and $g_{\rm \gamma N \Delta}$ is the coupling
constant, which we calculate below in our effective quark model by example
the mode $\Delta^+ \to p + \gamma$.

The computation of effective photon--hadron interactions in effective quark
models is a rather sensitive procedure demanding careful summation of all
possible diagrams containing both point--like photon--quark and
photon--hadron vertices. In order to simplify calculations we suggest to
replace the derivation of the ${\rm \gamma N \Delta}$--interaction by the
${\rm \rho^0 N \Delta}$--interaction, as photons in the decays $\Delta \to
N + \gamma$ are isovectors, and then to make a change $\rho^0_{\mu}(x) \to
(e/g_{\rho})\,A_{\mu}(x)$, where $g_{\rho}$ is the $\rho \pi \pi$--coupling
constant, according to the Vector Dominance hypothesis.

It is well--known [4,28,29] that the Vector Dominance hypothesis is
applicable not only to high energy processes but mainly at low energies to
the description of low--energy interactions of low--lying mesons within the
Effective Chiral Lagrangian approach [4,28]. For example, the amplitude of
the famous decay $\pi^0 \to \gamma + \gamma$ can be completely described
within the Vector Dominance hypothesis. Indeed, the amplitude of the $\pi^0
\to \gamma + \gamma$ decay defined by the Adler--Bell--Jackiw anomaly reads
[30]:
\begin{eqnarray*}
{\cal A}(\pi^0 \to \gamma \gamma) = - \frac{e^2}{8 \pi^2 F_0}.
\end{eqnarray*}
The same expression can be obtained via the intermediate $(\rho^0 +
\omega^0)$--state, i.e. $\pi^0 \to \rho^0 + \omega^0 \to \gamma + \gamma$,
by virtue of the direct transitions $\rho^0 \to \gamma$ and $\omega^0 \to
\gamma$ with the coupling constants $\rho^0_{\mu}(x) \to
(e/g_{\rho})\,A_{\mu}(x)$ and $\omega^0_{\mu}(x) \to
(e/3g_{\rho})\,A_{\mu}(x)$, respectively [28--30]:
\begin{eqnarray*}
{\cal A}(\pi^0 \to \gamma \gamma) = - g_{\pi
\rho\omega}\,\frac{e}{g_{\rho}}\,\frac{e}{3g_{\rho}} = -
\frac{3g_{\rho}^2}{8 \pi^2 F_0}\,\frac{e}{g_{\rho}}\,\frac{e}{3g_{\rho}} =
- \frac{e^2}{8 \pi^2 F_0}.
\end{eqnarray*}
Referring to this experience [4,28--30] the Vector Dominance hypothesis can
be readily applied to the description of the $\Delta^+ \to p + \gamma$
transition.

Since the $\rho^0$--meson couples to the quark current through the interaction
\begin{eqnarray}\label{label3.2}
{\cal L}^{\rm \rho^0 qq}(x) = \frac{g_{\rho}}{2}\,[\bar{u}(x)
\gamma^{\nu}u(x) - \bar{d}(x) \gamma^{\nu} d(x)]\,\rho^0_{\nu}(x),
\end{eqnarray}
the effective Lagrangian ${\cal L}^{\rm \rho^0 p \Delta^+ }_{\rm eff}(x)$
of the  $\Delta^+ \to p + \rho^0$ transition is defined by
\begin{eqnarray}\label{label3.3}
\hspace{-0.4in}&&\int d^4x {\cal L}^{\rm \rho^0 p \Delta^+}_{\rm eff}(x) =
- \Bigg(\frac{g_{\rho}}{2}\Bigg) \Bigg(\frac{g^2_{\rm
B}}{\sqrt{2}}\Bigg)\int d^4x d^4x_1 d^4x_2  \rho^0_{\nu}(x_2)\nonumber\\
\hspace{-0.4in}&&\bar{\Delta}^+_{\mu}(x)\,<0|{\rm T}(\eta^{\mu}_{\rm
\Delta^+}(x)[\bar{u}(x_2) \gamma^{\nu}u(x_2) - \bar{d}(x_2) \gamma^{\nu}
d(x_2)]\bar{\eta}_{\rm N}(x_1))|0> p(x_1),
\end{eqnarray}
where $\eta^{\mu}_{\rm \Delta^+}(x) =  \varepsilon^{ijk}\,[\bar{u^c}_i(x)
\gamma^{\mu} u_j(x)] d_k(x)$. By using the formulae of quark conversion
[12] (Ivanov) the matrix element in the r.h.s. of Eq.(\ref{label3.3}) can
be represented by the constituent quark diagrams and is given in terms of
the momentum integrals as follows:
\begin{eqnarray}\label{label3.4}
\hspace{-0.4in}&&\int d^4x\,{\cal L}^{\rm \rho^0 p \Delta^+}_{\rm eff}(x) =
\nonumber\\
\hspace{-0.5in}&&= -
\Bigg(\frac{g_{\rho}}{2\sqrt{2}}\Bigg)\Bigg(\frac{g_{\rm
B}}{8\pi^2}\Bigg)^2\int\! d^4x \!\int \frac{d^4x_1
d^4p_1}{(2\pi)^4}\frac{d^4x_2 d^4p_2}{(2\pi)^4}e^{ -i(x - x_1)\cdot
p_1}e^{-i(x - x_2)\cdot p_2}\rho^0_{\nu}(x_2)\nonumber\\
\hspace{-0.4in}&&\int \frac{d^4k_1 }{\pi^2i}\frac{d^4k_2}{\pi^2i}\,
\bar{\Delta}^+_{\mu}(x)\,\Bigg[\frac{1}{m - \hat{k}_1 - \hat{k}_2 -
\hat{p}_1 - \hat{p}_2}\gamma^{\nu} \frac{1}{m - \hat{k}_1 - \hat{k}_2 -
\hat{p}_1}\gamma_{\lambda}\gamma^5 \Bigg]  p(x_1) \,\nonumber\\
\hspace{-0.4in}&&{\rm tr}\Bigg\{\frac{1}{m +
\hat{k}_1}\gamma^{\mu}\frac{1}{m - \hat{k}_2}\gamma^{\lambda}\Bigg\} .
\end{eqnarray}
Since we use a cut--off regularization [12], the derivation of the gauge
invariant ${\rm \rho^0 N \Delta}$--interaction from the r.h.s. of
Eq.(\ref{label3.4}) is not straightforward procedure. In order to extract a
gauge invariant contribution we suggest to make an arbitrary shift of
virtual momenta proportional to the 4--momentum $p_2$ of the
$\rho^0$--meson. When applying then large $M_{\rm N}$ expansion and keeping
only linear terms in $\rho^0$--meson momentum expansion [12] one can fix
the parameter of the shift removing the gauge non--invariant contributions
[31]. The gauge invariant term turns out to be independent of the parameter
of the shift and reads
\begin{eqnarray}\label{label3.5}
\hspace{-0.4in}&&\int d^4x\,{\cal L}^{\rm \rho^0 p \Delta^+}_{\rm eff}(x) =
- \frac{3}{2}\Bigg(\frac{g_{\rho}}{\sqrt{2}}\Bigg) \Bigg(\frac{g_{\rm
B}}{3}\Bigg)^2\frac{<\bar{q}q>^2}{M^2_{\rm N}}\int\! d^4x \!\int
\frac{d^4x_1 d^4p_1}{(2\pi)^4}\frac{d^4x_2 d^4p_2}{(2\pi)^4} \nonumber\\
\hspace{-0.4in}&&e^{-i(x - x_1)\cdot p_1} e^{-i(x - x_2)\cdot
p_2}\,\bar{\Delta}^+_{\mu}(x) \Bigg(-\frac{2}{M_{\rm N}}\Bigg)
(p^{\mu}_2\gamma^{\nu} - \hat{p}_2 g^{\mu\nu}) \gamma^5 p(x_1)
\rho^0_{\nu}(x_2).
\end{eqnarray}
This yields the effective Lagrangian ${\cal L}^{\rm \rho^0 p \Delta^+}_{\rm
eff}(x)$:
\begin{eqnarray}\label{label3.6}
{\cal L}^{\rm \rho^0 p \Delta^+}_{\rm eff}(x) = i\,\frac{g_{\rho}}{2M_{\rm
N}}\,\Bigg[\frac{\sqrt{2}}{3}\,g^2_{\rm B}\frac{<\bar{q}q>^2}{M^2_{\rm
N}}\Bigg]\bar{\Delta}^+_{\mu}(x)\,\gamma_{\nu}\gamma^5 p(x)\,{\cal
F}^{\mu\nu}(x) + {\rm h.c.},
\end{eqnarray}
where ${\cal F}_{\mu\nu}(x) = \partial_{\mu}\rho^0_{\nu}(x) -
\partial_{\nu}\rho^0_{\mu}(x)$. Making then a shift $\rho^0_{\mu}(x) \to
(e/g_{\rho})\,A_{\mu}(x)$ caused by the Vector Dominance hypothesis we
derive the effective Lagrangian of the ${\rm \gamma N \Delta}$--interaction
\begin{eqnarray}\label{label3.7}
{\cal L}^{\rm \gamma p \Delta^+}_{\rm eff}(x) = i\,\frac{e}{2M_{\rm
N}}\,\Bigg[\frac{\sqrt{2}}{3}\,g^2_{\rm B}\frac{<\bar{q}q>^2}{M^2_{\rm
N}}\Bigg]\bar{\Delta}^+_{\mu}(x)\,\gamma_{\nu}\gamma^5 p(x)\,F^{\mu\nu}(x)
+ {\rm h.c.}
\end{eqnarray}
Thus, the coupling constant $g_{\rm \gamma N \Delta}$ is given by
$g_{\rm \gamma N \Delta}/g_{\rm \pi N \Delta} = (\sqrt{3}/4)(F_0/m) =
0.12$. This ratio agrees  with the $SU(6)$ prediction $g_{\rm \gamma N
\Delta}/g_{\rm \pi N\Delta} = 0.14$ [32]. The agreement of our result
$g_{\rm \gamma N \Delta}/g_{\rm \pi N \Delta} = 0.12$ with other
phenomenological approaches [26,27] is only qualitative.

\section{$\sigma_{\rm \pi N}$--term}
\setcounter{equation}{0}

In this section we calculate current quark mass corrections to the mass of
the proton $M_{\rm p}$ and the $\sigma_{\rm \pi N}$--term, defined as
\begin{eqnarray}\label{label4.1}
\sigma_{\rm \pi N} = m_0\,<p|[\bar{u}(0) u(0) + \bar{d}(0) d(0)]|p>
\end{eqnarray}
and related to these chiral corrections through the Feynman--Hellmann theorem
\begin{eqnarray}\label{label4.2}
\sigma_{\rm \pi N} = m_0\,\frac{\partial M_{\rm p}(m_0)}{\partial m_0},
\end{eqnarray}
where $m_0 = (m_{0 u} + m_{0 d})/2 = 5.5\,{\rm MeV}$ is an averaged current
quark mass at $m_{0 u} = 4\,{\rm MeV}$ and $m_{0 d} = 7\,{\rm MeV}$ [33].
The calculation of the $\sigma_{\rm \pi N}$--term has a long history [34].
The current magnitudes of $\sigma_{\rm \pi N}$ obtained on lattice are
$40\div 60\,{\rm MeV}$ [35] and 50$\,{\rm MeV}$ [36].

In our effective quark model current quark mass corrections to the mass of
the proton can be defined by the effective Lagrangian
\begin{eqnarray}\label{label4.3}
\hspace{-0.4in}&&\int d^4x\,\delta{\cal L}^{\rm pp}_{\rm eff}(x) =
m_0\,\Bigg(\frac{g^2_{\rm B}}{2}\Bigg) \,\int\  d^4x d^4x_1
d^4x_2\nonumber\\
\hspace{-0.4in}&&\bar{p}(x) <0|{\rm T}(\eta_{\rm N}(x)[\bar{u}(x_2) u(x_2)
+ \bar{d}(x_2) d(x_2)]\bar{\eta}_{\rm N}(x_1))|0>p(x_1).
\end{eqnarray}
By using the formulae of quark conversion [12] (Ivanov) the matrix element
in the r.h.s. of  Eq.(\ref{label4.3}) can be represented by the constituent
quark diagrams and is given in terms of the momentum integrals as follows:
\begin{eqnarray}\label{label4.4}
\hspace{-0.4in}&&\int d^4x\,\delta{\cal L}^{\rm pp}_{\rm eff}(x) =
m_0\,\frac{3}{2}\,\Bigg(\frac{g_{\rm B}}{8\pi^2}\Bigg)^2 \,\int d^4x \int
\frac{d^4x_1d^4p_1}{(2\pi)^4} e^{ -i(x - x_1)\cdot p_1}\nonumber\\
\hspace{-0.4in}&&\int
\frac{d^4k_1}{\pi^2i}\frac{d^4k_2}{\pi^2i}\,\bar{p}(x)
\Bigg(\frac{\bar{v}}{4m}\Bigg)\Bigg[ \gamma_{\mu}\gamma^5 \frac{1}{m -
\hat{k}_1 - \hat{k}_2 -\hat{p}_1}\frac{1}{m - \hat{k}_1 - \hat{k}_2
-\hat{p}_1} \gamma_{\nu}\gamma^5\Bigg] p(x_1)\nonumber\\
\hspace{-0.4in}&&{\rm tr}\Bigg\{\frac{1}{m +
\hat{k}_1}\gamma^{\nu}\frac{1}{m - \hat{k}_2}\gamma^{\mu}\Bigg\} +
m_0\,\frac{3}{4}\,\Bigg(\frac{g_{\rm B}}{8\pi^2}\Bigg)^2 \,\int d^4x \int
\frac{d^4x_1d^4p_1}{(2\pi)^4} e^{ -i(x - x_1)\cdot p_1}\nonumber\\
\hspace{-0.4in}&&\int
\frac{d^4k_1}{\pi^2i}\frac{d^4k_2}{\pi^2i}\,\bar{p}(x) \Bigg[
\gamma_{\mu}\gamma^5 \frac{1}{m - \hat{k}_1 - \hat{k}_2 -\hat{p}_1}
\gamma_{\nu}\gamma^5\Bigg] p(x_1)\nonumber\\
\hspace{-0.4in}&&\Bigg(\frac{\bar{v}}{4m}\Bigg){\rm tr}\Bigg\{\frac{1}{m +
\hat{k}_1}\frac{1}{m + \hat{k}_1}\gamma^{\nu}\frac{1}{m -
\hat{k}_2}\gamma^{\mu} + \frac{1}{m + \hat{k}_1}\gamma^{\nu}\frac{1}{m -
\hat{k}_2}\frac{1}{m - \hat{k}_2}\gamma^{\mu}\Bigg\},
\end{eqnarray}
where $\bar{v} = - <\bar{q}q>/F^2_0 =1.92\,{\rm GeV}$ [12] and the factor
$\bar{v}/4m$ appears due to the contribution of the exchange of the
isoscalar scalar $\sigma$--meson with the quark structure $(\bar{u}u +
\bar{d}d)/\sqrt{2}$ and the mass $M_{\sigma} = 2m$ [12,37]. The calculation
of the integrals entering Eq.(\ref{label4.4}) is analogous to that having
been performed in preceding sections. The resultant effective Lagrangian
$\delta {\cal L}^{\rm pp}_{\rm eff}(x)$ is given by
\begin{eqnarray}\label{label4.5}
\hspace{-0.4in}&&\delta{\cal L}^{\rm pp}_{\rm eff}(x) =\nonumber\\
\hspace{-0.4in}&&=- \Bigg[m_0 \frac{2}{3} g^2_{\rm B}
\frac{<\bar{q}q>^2}{M^2_{\rm N}}\Bigg(\frac{\bar{v}}{4m}\Bigg) + m_0
\frac{3}{4} \frac{M_{\rm N}}{m} g^2_{\rm B}\,\frac{<\bar{q}q>^2}{M^2_{\rm
N}}\Bigg(\frac{\bar{v}}{4m} - 1\Bigg)\Bigg]\,\bar{p}(x) p(x) = \nonumber\\
\hspace{-0.4in}&&=-\Bigg[m_0 \,g_{\rm \pi NN}
\frac{F_0}{m}\Bigg(\frac{\bar{v}}{4m}\Bigg) + m_0\,g_{\rm \pi
NN}\frac{9}{8}\,\frac{F_0}{m}\frac{M_{\rm N}}{m}\,\Bigg(\frac{\bar{v}}{4m}
- 1\Bigg)\Bigg]\,\bar{p}(x) p(x).
\end{eqnarray}
This yields the proton mass as a function of $m_0$:
\begin{eqnarray}\label{label4.6}
M_{\rm p}(m_0) = M_{\rm N} + m_0\,g_{\rm \pi
NN}\,\frac{F_0}{m}\Bigg(\frac{\bar{v}}{4m}\Bigg)\Bigg[1 +
\frac{9}{8}\,\frac{M_{\rm N}}{m}\,\Bigg(1 - \frac{4m}{\bar{v}}\Bigg)\Bigg].
\end{eqnarray}
Due to Eq.(\ref{label4.2}) we arrive at the $\sigma_{\rm \pi N}$--term:
\begin{eqnarray}\label{label4.7}
\sigma_{\rm \pi N} =  m_0\,g_{\rm \pi
NN}\,\frac{F_0}{m}\Bigg(\frac{\bar{v}}{4m}\Bigg)\Bigg[1 +
\frac{9}{8}\,\frac{M_{\rm N}}{m}\,\Bigg(1 - \frac{4m}{\bar{v}}\Bigg)\Bigg]
= 60\,{\rm MeV}.
\end{eqnarray}
The numerical value of the $\sigma_{\rm \pi N}$--term is obtained at
$M_{\rm N}= 940\,{\rm MeV}$ and $g_{\rm \pi NN} =13.4$. Our estimate
$\sigma_{\rm \pi N}=60\,{\rm MeV}$ agrees with the numerical data obtained
on lattice [35,36].

\section{Conclusion}

We have suggested an effective quark model for octet and decuplet of
low--lying baryons. This model is some kind of baryon extension of the
ENJL--model with linear realization of chiral $U(3)\times U(3)$ symmetry
[12].  Baryons are included as external heavy states coupled to the
three--quark currents the spinorial structure of which is fixed within QCD
with linearly rising interquark potential [14] incorporating the
ENJL--model [13]. As has been shown in [13] the mass spectra of low--lying
mesons in QCD with linearly rising confinement potential coincide with the
mass spectra of mesons in the ENJL model. Then, the constituent quark mass
$m$ can be expressed in terms of the string tension $\sigma$ as follows $m
= 2\sqrt{\sigma}/\pi$. At $\sigma \sim 440\,{\rm MeV}$ we get $m \sim
300\,{\rm MeV}$. For the description of low--energy dynamics of baryons we
have added two new low--energy parameters with respect to the ENJL--model.
These are the phenomenological coupling constant $g_{\rm B}$ and the
averaged mass of the octet $M_{\rm N}$. The averaged mass of the decuplet
$M_{\rm \Delta}$ can be kept of order $M_{\rm N}$. In terms of these
parameters and the parameters of the ENJL--model, fixed by the low--energy
phenomenology of low--lying mesons [12], we have described the ${\rm \pi
NN}$, ${\rm \pi N \Delta}$ and ${\rm \gamma N \Delta}$ interactions and the
$\sigma_{\rm \pi N}$--term related to the amplitude of the elastic
low--energy ${\rm \pi N}$--scattering. The obtained results $g_{\rm \pi N
\Delta}/g_{\rm \pi NN} = 2$, $g_{\rm \gamma N \Delta}/g_{\rm \pi N \Delta}
= 0.12$ and $\sigma_{\rm \pi N} = 60\,{\rm MeV}$ agree reasonably well with
the data on low--energy phenomenology of baryons obtained both
experimentally and within other effective field theory approaches.

Concluding the discussion we would like to make the relation of our
approach to QCD much more obvious. It is well-known that QCD at low
energies should realize the main non--perturbative phenomena such as
hadronization, spontaneous breaking of chiral symmetry (SB$\chi$S) and
confinement. At present there is a consensus that the quark confinement
realizes itself through a linearly rising interquark potential. As has been
shown in Ref.[13] a linearly rising interquark potential is also
responsible for SB$\chi$S. The formation of a linearly rising interquark
potential is caused by the colour electric flux induced by the low--energy
gluon field configurations. Integrating out gluon degrees of freedom around
the low--energy gluon field configurations providing the formation of a
linearly rising interquark potential one should arrive at an effective
theory, an effective low--energy QCD, containing only quark degrees of
freedom. The quark system described by this effective theory  possesses
chirally invariant and chirally broken phases. The transition to the
chirally broken phase, i.e. the phenomenon of SB$\chi$S, accompanies itself
the hadronization, i.e. bosonization and baryonization, describing the
appearance of quark bound states with quantum numbers of low--lying mesons
$q\bar{q}$, baryons $qqq$ and so on. Due to the quark confinement all
observed quark bound states should be colourless. As in such an effective
low--energy QCD the gluon degrees of freedom have been integrated out, all
low--energy interactions of hadrons should be defined in terms of
quark--loop diagrams.

Since nowadays in continuum space--time formulation of QCD the
integration over gluon degrees of freedom can be hardly
carried out explicitly, the approximate schemes of this
integration admitting an analytical investigation are rather actual and
play an important role for the understanding of the behaviour of the
effective QCD at low energies. In our effective quark model
with chiral $U(3)\times U(3)$ symmetry the result of the
integration over gluon degrees of freedom concerning bosonization and
baryonization is represented by the effective Lagrangian incorporating the
effective local four--quark interactions responsible for SB$\chi$S, the
linearalized version of which contains quark--meson interactions like
Eq.(\ref{label2.2}) and Eq.(\ref{label3.2}), and the quark--baryon
interactions like Eq.(\ref{label1.3}). The former
corresponds to the phenomenological description of the appearance of the
tree--quark bound states -- baryons. We emphasize that the spinorial
structure of the three--quark currents $\eta_{\rm B_{\underline{8}}}(x)$
and $\eta_{\rm B_{\underline{10}}}(x)$ of the baryon octet and decuplet,
respectively, is not unambiguously defined if we would follow only the
$SU(3)$ or $SU(3) \times SU(2)_{\rm spin} \to SU(6)$ flavour symmetry.
Indeed, the three--quark currents with quantum numbers of the baryon octet
and decuplet can be constructed from the scalar, pseudoscalar, vector and
axial--vector diquark densities transforming like
$(\tilde{\underline{3}}_{\,\rm f},\tilde{\underline{3}}_{\,\rm c})$ and
$({\underline{6}}_{\,\rm f},\tilde{\underline{3}}_{\,\rm c})$,
respectively. The general form of the three--quark currents should contain
the contributions of these diquark densities with four arbitrary coupling
constants. However, as has been shown in Ref.[13], due to a linearly rising
interquark potential, i.e. a dynamics induced by  low--energy gluon field
configurations, the spinorial structure of the three--quark currents with
quantum numbers of the baryon octet and decuplet is fixed unambiguously and
defined by Eq.(\ref{label1.1}) and Eq.(\ref{label1.2}). Thus, the structure
of the three--quark currents, which we have used in the approach, is
completely caused by the low--energy properties of QCD leading to the quark
confinement.

The phenomenological character of the result of the integration over
low--energy gluon configurations, responsible for the formation of a
linearly rising interquark potential and leading to the baryonization of
the effective low--energy QCD, is accentuated by the inclusion of the
phenomenological coupling constant $g_{\rm B}$ which is an analogy of the
quark--meson coupling constants $g_{\rm \pi qq}$ and $g_{\rho}$ describing
phenomenologically the bosonization of QCD. The coupling constants $g_{\rm
B}$, $g_{\rm \pi qq}$ and $g_{\rho}$ define the vertices of low--energy
meson-baryon interactions in terms of the constituent quark--loop diagrams,
the virtual momenta of which are restricted by the SB$\chi$S scale
$\Lambda_{\chi} = 0.94\,{\rm GeV}$. As has turned out the effective
coupling constants of the ${\rm \pi NN}$, ${\rm \pi N\Delta}$ and ${\rm
\gamma N\Delta}$ interactions depend on $g_{\rm B}$ and the quark
condensate being a quantitative measure of SB$\chi$S in the effective
low--energy QCD.

Therefore, the spinorial structure of the three--quark currents with
quantum numbers of the baryon octet and decuplet and the proportionality of
the effective coupling constants of the meson(photon)--baryon--baryon
interactions to the quark condensate testify the direct relation of all
low--energy interactions of the baryon octet and decuplet to
non--perturbative phenomena of low--energy QCD. Thus, in our effective
quark model with chiral $U(3)\times U(3)$ symmetry the existence and the
low--energy interactions of the observed baryon octet and decuplet are
completely caused by low--energy non-perturbative properties of QCD leading
to the quark confinement.

\section{Acknowledgment}

This work was supported in part by the Slovak Grant Agency for Science,
Grant No. 2/4111/97 (M.Nagy).

\end{document}